\documentclass[preprint,dvipdfmx]{ptephy_v1}
\usepackage{graphicx}
\usepackage{bm}
\usepackage{color}
\newcommand{\nn}{\nonumber}
\newcommand{\be}{\begin{equation}}
\newcommand{\ee}{\end{equation}}
\newcommand{\bea}{\begin{eqnarray}}
\newcommand{\eea}{\end{eqnarray}}
\newcommand{\bean}{\begin{eqnarray*}}
\newcommand{\eean}{\end{eqnarray*}}
\newcommand{\bit}{\begin{itemize}}
\newcommand{\eit}{\end{itemize}}
\newcommand{\ben}{\begin{enumerate}}
\newcommand{\een}{\end{enumerate}}
\def\ran{\rangle}
\def\lan{\langle}
\def\bra #1{\lan #1 |}
\def\ket #1{| #1 \ran}

\makeatletter
\renewcommand\ps@headings{\let\@mkboth\markboth
\renewcommand\@oddfoot{\hbox to \textwidth{\hfil\rmfamily\thepage\hfil%
\makebox[0\p@][r]{\@typeset}}}}
\makeatother

\begin{document}

\title{
Nonperturbative evaluation\\of quantum 
particle production\\in parametric resonance enhanced by noise
}
\author{Asako Murakami$^*$ and Haruhiko Terao}
\affil{
Department of Physics, Nara Women's University
Nara, 630-8506, Japan \email{amurakami@asuka.phys.nara-wu.ac.jp}
}%

\begin{abstract}
 Numerical studies are reported to support the idea to
 explain the particle production in late inflationary era based on the
 parametric resonance with a noise effect.  Two nonperturbative
 renormalization group formulations are used to numerically calculate
 the time evolution of particle numbers.  Firstly, the dynamical
 renormalization group (DRG) method is applied to sum up the secular
 contributions.  Secondly, we derive an exact evolution equation of the
 particle number which turned out to be possible surprisingly owing to
 the presence of the noise effect.  Our numerical results show a drastic
 enhancement of the particle production in agreement with earlier works
 qualitatively.  A comparison is made of numerical results based on two
 methods.  Our work provides nonperturbative and quantitative methods to
 evaluate the particle production for the inflationary cosmology. 
\end{abstract}

\subjectindex{B32,B72,B76}

\maketitle

\section{Introduction}

Recent observations on the cosmic microwave background offer us strong
evidence for the inflation at the beginning of the universe. Explosive
particle production at the end of the inflation era is a crucial
aspect of inflationary cosmology in order to realize the starting point
of the successful big bang universe. It is supposed that a coherent
oscillation of a scalar field, called the inflaton field, at the end of
inflation period transfers the energy to matter particles very rapidly.
Explicitly, an oscillation of the inflaton field induces a parametric
resonance instability for any bosonic modes coupled to the field and
leads to the explosive particle
production~\cite{Kofman:1994rk,Kofman:1997yn,Traschen:1990sw,
Shtanov:1994ce}.

The particle production by the parametric resonance has been discussed in
many papers. The most simplified model is a harmonic oscillator
receiving an external periodic force: its classical equation of
motion is nothing but the Mathieu equation. While parametric resonance
is a typical non-equilibrium phenomena, perturbative approximation is
known to break down due to presence of the secular contributions in
general non-equilibrium quantum systems. Therefore, nonperturbative
analyses are required in order to evaluate parametric resonance quantum
mechanically. 

Among nonperturbative methods, those based on the renormalization group
offers us powerful schemes for various dynamical
problems~\cite{WilsonRG}.

The dynamical renormalization group (DRG) is
frequently applied to pursue a dynamical evolution and to sum up the
secular contributions. This corresponds to the Gell-Mann Low
renormalization group in quantum field theories based on the
perturbative expansion~\cite{Boyanovsky:1998aa,Boyanovsky:1999cy}.

The exact renormalization group has also been a
powerful tool for the non-perturbative analyses. Recently the functional
renormalization group for the non-equilibrium field theories are
proposed~\cite{Gasenzer:2008zz,Gasenzer:2010rq}. There a cutoff is
introduced to the time of the propagators and the cutoff dependence of
the 1PI functionals are derived. The cutoff time dependence is, of course,
compatible with the time evolution by the equation of motion.

We found that the particle production in the parametric resonance can
be analyzed with the functional renormalization group approach by
introducing the cutoff time to the external forces \cite{HMNT}.
It is shown that the particle number can be traced exactly
by using the functional renormalization group equations.
In this paper, 
we apply the method of Ref.~\cite{HMNT} to our model
to be given in the next section.

There are the resonant regions and the oscillating regions in the
particle number spectrum. This just corresponds to the unstable regions
and the stable regions in the parameter space of the Mathieu equation.
Brandenberger et. al. pointed out that the stable region of the Mathieu
equation disappears in the presence of noise
\cite{Zanchin:1997gf,Brandenberger:2014mba}.
%
%
Moreover, it has been proven that the production rate of the particle
number, i.e., the Floquet exponent, is always enhanced irrespectively of
the particle momentum.  However an explicit way to evaluate the mean
value of the particle number was not given there.

In this paper, we treat the non-equilibrium dynamics of the particle
production in the real scalar quantum field theory.  We note that the
noise effect can be replaced by a self-interaction term with an
imaginary coupling constant after taking the average over the various
sample of noise functions.  We formulate the renormalization equations
for this system.  Then the dynamical renormalization shows explicitly
the disappearance of stable regions as well as the enhancement of the
particle production rate.  It is found, surprisingly, that the time
evolution of the particle number can be traced exactly owing to the
closed time path structure.  We also study this exact
evolution equation numerically.

In Section 2 our model of a real scalar field is explained. There
the noise effect is given as a self-interaction.  In Section 3 we
consider the DRG for the particle numbers.  In Section 4 we derive the
exact evolution equation for the particle number, which may be regarded
as the exact renormalization group equations.  Section 5 is devoted to
the numerical analyses of these renormalization group equations.  There
it will be shown that the enhancement of the particle number by the
noise effect certainly occurs as expected.  The last section is for the
conclusion and the discussions.

We added two appendices to discuss the noise effect on the Mathieu
equation.  Appendix~A recapitulates an important result of
\cite{Zanchin:1997gf}: a solution grows exponentially even in stable
regions of parameters.  We propose a noble simulation method in
Appendix~B to observe the drastic growth of the particle number in this
classical setting.

\section{Model for the parametric resonance with noise}

\subsection{Model for the parametric resonance}

We consider the following model for the parametric resonance in the
environment of noise; the Lagrangian density is given by
\be
{\cal L}= \frac{1}{2}\partial_{\mu}\chi \partial^{\mu}\chi
-\frac{1}{2}m_{\chi}^2 \chi^2 -\frac{1}{2} \kappa \phi \chi^2
-\frac{1}{2}g q(t) \chi^2,
\ee
where $\chi$ denotes a real scalar field and $q(t)$ represents the
external homogeneous noise. 
The field $\phi$ is  the inflaton field and oscillates in the end of the
inflation 
period. We simply treat this as a homogeneous classical background given
as $\phi(t)=\Phi \cos m_{\phi} t$. We do not take account of expansion
of the universe and use the stationary flat metric. 

When both of the external force and the noise are homogeneous, the model
is reduced to an assembly of quantum mechanical systems by decomposing
the scalar field as, $\chi({\bm x},t)=\sum_{\bm k} \chi_{\bm k}(t) e^{i
{\bm k}{\bm x}}$.  Then the Hamiltonian is given by
\be H= \sum_{\bm k}
\frac{1}{2}|\dot{\chi}_{\bm k}(t)|^2 +\frac{1}{2}\left( \omega_k^2 +
\kappa \Phi \cos m_{\phi}t + gq(t)\right)|\chi_{\bm k}(t)|^2, \ee
where
$\omega_k^2 = {\bm k}^2 + m_{\chi}^2$.  We also define $m_{\phi}=\omega_0$,
$\kappa \Phi=2 \epsilon \omega_0^2$, $\omega_k^2 = \omega_0^2 a_k$.
Then the equation of motion is given by \be \ddot{\chi}_{\bm k} +
\omega_0^2(a_k + 2 \epsilon \cos \omega_0 t + g q(t)) \chi_{\bm k} = 0.
\ee

Without the noise term this equation is known as the Mathieu equation.
It is well-known that the resonant region and the stable region forms a
band structure in the parameter space of $(\epsilon, a_k)$.  In the
resonance band the particle number increases exponentially, while it
just oscillates in the stable band.  The resonance occurs when
$\omega_k^2 \simeq (n/2)^2 \omega_0^2, (n=1,2,3 \cdots)$ for a small
$\epsilon$.  However it has been shown that in
Ref.~\cite{Zanchin:1997gf} that the particle number increases
exponentially irrespectively of the particle momentum $\bm k$ in the
presence of the noise, and therefore the stable band disappears.

In Appendices, we discuss the solutions of the equation motion given by
(3).  The solution of the Mathieu equation is known to enjoy the
characteristic form as \be \chi_{\bm k}(t) = P_{\bm k}(t)e^{\gamma_{\bm
k}t}, \ee where $P_{\bm k}(t)$ is a periodic function. The factor
$\gamma_{\bm k}$ is called the Floquet exponent.  When this exponent has
a real positive part, then the resonance occurs.
In~Ref.\cite{Zanchin:1997gf}, it was shown that real part of the
exponent is always enhanced in the presence of the noise. In Appendix~A,
we briefly sketch their argument.
The classically defined particle number is evaluated numerically using
the equation of motion (3) in Appendix~B.

We may evaluate the particle number by perturbative expansion around the first
resonant peak, $\omega=\omega_0/2$.
If we parametrize as
\be
a_k=\frac{1}{4}+\epsilon \tilde{a}_k,
\ee
then the Hamiltonian may be decomposed to the free part and the interaction part as
\bea
H&=& \sum_{\bm k} \left\{ 
\frac{1}{2}|\dot{\chi}_{\bm k}(t)|^2+
\frac{1}{2} \omega^2 |\chi_{\bm k}(t)|^2\right\} + H'+H_q, \\
H'&=& \sum_{\bm k} \frac{1}{2} \epsilon \omega_0^2 \left(\tilde{a}_k + 2 \cos \omega_0 t \right)
|\chi_{\bm k}(t)|^2
\equiv  \sum_{\bm k} \frac{1}{2} \epsilon f(t)|\chi_{\bm k}(t)|^2, \label{externalforce} \\
H_q &=&  \sum_{\bm k} \frac{1}{2} g q(t)|\chi_{\bm k}(t)|^2.  \label{noiseterm}
\eea

In order to evaluate the particle number we introduce the particle creation and annihilation
operators.
\be
\chi_{\bm k}(t)=\frac{1}{\sqrt{2 \omega}}(a_{\bm k}(t)+a_{\bm k}^{\dagger}(t))
\ee
Then the particle number is given by
\be
n_{\bm k}(t) = 
\mbox{Tr}\left\{ \rho(0) a_{\bm k}^{\dagger}(t) a_{\bm k}(t) \right\} 
\equiv \lan a_{\bm k}^{\dagger}(t) a_{\bm k}(t) \ran,
\ee
where $\rho(0)$ denotes the initial density matrix at $t=0$. We assume that the initial
state is given by the vacuum state $\ket 0$, {\rm i.e}, $\rho(0)=\ket 0 \bra 0 $.
Then
\be
n_{\bm k}(t)=\bra 0 a_{\bm k}^{\dagger}(t) a_{\bm k}(t) \ket 0
\ee
  
  We also define the anomalous particle numbers as
\bea m_{\bm k}(t) &=&
\bra 0 a_{\bm k}(t) a_{\bm k}(t) \ket 0, \\ m_{\bm k}^*(t) &=& \bra 0
a_{\bm k}^{\dagger}(t) a_{\bm k}^{\dagger}(t) \ket 0, \eea for the later
conveniences.  Similarly the two point functions are defined by \be
G_{\bm k}(t,t') = \langle T_c \chi_{\bm k}(t) \chi_{\bm k} (t') \rangle,
\ee where $T_c$ denotes the time ordering along the closed time path.
It will be shown later that both the particle numbers and the anomalous
particle numbers are derived from the two point functions.

\subsection{Statistical average of the white noise}

Our present interest is the statistical average of the particle numbers
obtained under an assembly of noise function $q$.  We assume here that
the noise is given by the Gaussian white noise such that \be \langle
q(t) q(t') \rangle = \sigma^2 \delta(t-t').  \ee Such noise could be
caused by quantum fluctuations of the inflaton field or of various other
environmental fields coupled to $\chi$, or could be some thermal
fluctuations.  In this paper we do not specify the origin of the noise
and simply treat the parameters $g$ and $\sigma$ as free parameters.

Then the two point function after taking the statistical average of the noise is
given by
\bea
G_{\bm k} (t,t')
&=& \langle T_c \chi_{\bm k}(t) \chi_{\bm k}(t') \rangle \nn \\
&=& \int Dq e^{-\frac{1}{2\sigma^2} \int dt q(t)^2}
 \langle T_c \chi_{\bm k}(t) \chi_{\bm k}(t') \rangle_{q},
\eea
where $\langle ~~~\rangle_q$ denotes the correlation obtained under a fixed noise
function $q(t)$.

The path integral representation for the two point function is useful.
By the Keldish-Schwinger formalism, the path integral is given on the closed time path.
Explicitly the two point function is given by  
\bea
G_{\bm k} (t,t')
&=&
\int D\chi^+ D\chi^- \chi^+_{\bm k}(t) \chi^-_{\bm k}(t')~
e^{
i \int dt \left({\cal L}_0[\chi^+]-{\cal L}_0[\chi^-]\right)
} \nn \\
& &
\times \int Dq~ e^{-\frac{1}{2\sigma^2} \int dt q(t)^2}
e^{
-\frac{i}{2} g \int dt q(t) 
\sum_{\bm k} \left( \chi^+_{\bm k}(t)^2 - \chi^-_{\bm k} (t)^2 \right)
} \\
&=&
 \int D\chi^+ D\chi^- \chi^+_{\bm k}(t) \chi^-_{\bm k}(t')~
e^{i S_{\rm eff}[\chi^+, \chi^-]},
\eea
where ${\cal L}_0$ denotes the Lagrangian density for $q=0$.
The ``effective action'' $ S_{\rm eff}$ is given explicitly by
\bea
S_{\rm eff}&=&
\int dt \Bigl\{
{\cal L}_0[\chi^+]-{\cal L}_0[\chi^-]  \nn \\
& &
\hspace{24pt} 
+ i \frac{\sigma^2 g^2}{8}  \sum_{\bm k} \sum_{{\bm k}'} 
\left( \chi^+_{\bm k}(t)^2 - \chi_{\bm k}^- (t)^2 \right)
\left( \chi^+_{{\bm k}'}(t)^2 - \chi_{{\bm k}'}^- (t)^2 \right)
\Bigr\}.
\eea

It should be noticed that there appear self-interaction terms with an
imaginary coupling constant.  Moreover, it is remarkable that the $(+)$
fields and the $(-)$ fields interact each other through these terms.
Actually, it has been shown that such interaction appears generally in
the Wilsonian effective action for a non-equilibrium field theory
\cite{Gasenzer:2008zz}\cite{Gasenzer:2010rq}.  After integrating the
quantum fluctuations with high frequencies, such stochastic noise terms
as well as the dissipative terms appear for the lower frequency modes.
Thus it would be more natural to treat such self-interaction terms from
the beginning.

\section{Dynamical renormalization group}

We first perform a perturbative expansion to evaluate the two point
function. Suppose the density matrix is known at an arbitrary time
$t=\tau$. Namely we assume the particle numbers as well as the anomalous
particle numbers are given at $t=\tau$. Then the two point function for
a free particles are found to be 
\bea
D^>_{\bm k}(t,t')
&=& \langle \chi_{\bm k}(t) \chi_{\bm k}(t') \rangle \nn \\
&=& 
\frac{1}{2 \omega} \left\{
 e^{-i \omega(t-t')}\left(n_{\bm k}(\tau)+1 \right) \
 +e^{i \omega(t-t')}n_{\bm k}(\tau)  \right. \nn \\
& & 
\hspace{24pt}
\left.
+ e^{-i \omega(t+t')}\tilde{m}_{\bm k}(\tau)
+ e^{i \omega(t+t')}\tilde{m}_{\bm k}^*(\tau)
\right\}\,.
\eea
Here we used the free particle evolution, 
$a_{\bm k}(t) = \exp[-i \omega(t-\tau)] a_{\bm k}(\tau)$,
and the rescaling of the anomalous particle numbers, $\tilde{m}_{\bm
k}(\tau)=e^{2i\omega \tau}m_{\bm k}(\tau)$.

In the first order of the perturbative expansion, the two point function
is obtained as
\bea
G^>_{\bm k}(t,t') &=&  
D_{\bm k}^>(t,t') -i\epsilon \int_{\tau} ds~ f(s)
\left[
 D_{\bm k}^>(t,s) D_{\bm k}^>(t',s) -D_{\bm k}^>(s,t) D_{\bm k}^>(s,t')
\right] \nn \\
& &
-g^2 \frac{\sigma^2}{8}\int_{\tau} ds~ \Bigl\{
 D_{\bm k}^>(t,s) D_{\bm k}^>(t',s) 
+  D_{\bm k}^>(s,t) D_{\bm k}^>(s,t') \nn \\
& &
\hspace{60pt}
- D_{\bm k}^>(t,s) D_{\bm k}^>(s,t') 
- D_{\bm k}^>(s,t) D_{\bm k}^>(t',s)
\Bigr\} D_{\bm k}^>(s,s) \nn \\
&=&
D_{\bm k}^>(t,t') -\epsilon \int_{\tau} ds~ f(s)
\left[
 D_{\bm k}^>(t',s) \rho_{\bm k}(t-s) + D_{\bm k}^>(s,t) \rho_{\bm k}(t'-s)
\right] \nn \\
& &
+ g^2 \frac{\sigma^2}{8}\int_{\tau} ds~
\rho_{\bm k}(t-s)\rho_{\bm k}(t'-s) D_{\bm k}^>(s,s).
\eea
Here we introduced the spectral function for the free field as
\bea
\rho_{\bm k}(t-s) &=& 
i\left( D_{\bm k}^>(t,s)-D_{\bm k}^>(s,t)  \right) \nn \\
&=& \frac{1}{\omega} \sin[\omega(t-s)].
\label{freerho}
\eea

Then we may perform the time integration by using the explicit form for
the external force (\ref{externalforce}) and  the two point function is
found to be
\bea
G^>_{\bm k}(t,t') 
&=&  D_{\bm k}^>(t,t') \nn \\
& &
+ i \epsilon (t-\tau) \Bigl\{ 
2\cos \omega(t-t') \left(\tilde{m}_{\bm k}(\tau)-\tilde{m}^*_{\bm k}(\tau) \right)   \nn \\
& &\hspace{60pt}
+ e^{i\omega(t+t')}\left( 2 \tilde{a}_{\bm k} \tilde{m}^*_{\bm k}(\tau) + 2 n_{\bm k}(\tau) +1 \right)
\nn \\
& &\hspace{60pt}
- e^{-i\omega(t+t')}\left( 2 \tilde{a}_{\bm k} \tilde{m}_{\bm k}(\tau) + 2 n_{\bm k}(\tau) +1 \right)
\Bigr\} \nn \\
& &
+ g^2\frac{\sigma^2}{64\omega^3}(t-\tau)
\Bigl\{
2\cos[\omega(t-t')]\left(2 n_{\bm k}(\tau) + 1 \right) \nn \\
& &
\hspace{72pt}
-e^{i\omega(t+t')}\tilde{m}^*_{\bm k}(\tau)
-e^{-i\omega(t+t')}\tilde{m}_{\bm k}(\tau) \Bigr\} \nn \\
& &
+ \mbox{(non-secular terms)}.
\eea
The contributions proportional to $t-\tau$ are called the secular
terms. Other non-secular terms are  oscillating and small
comparatively. Therefore, these are neglected in applying the DRG. It is seen that the perturbative expansion breaks
down for $\epsilon(t-\tau) > 1$ due to the secular terms.
The DRG is a simple method to sum up the
secular contributions in all orders.   

Now we derive the DRG equations for the particle
numbers. The key point is that the two point function should be
independent of choice of the ``renormalization point''
$\tau$.\footnote{This idea is corresponding with the Callan-Symanzik
equation in the perturbation theory of the quantum field theories. }
Explicitly we impose 
\be
\frac{d }{d \tau}G_{\bm k}^>(t,t') = 0.
\ee
Then we may find $\tau$ dependence of $n_{\bm k}(\tau)$ 
as well as $\tilde{m}_{\bm k}(\tau)$ up to $O(\epsilon^2, g^4)$ as
\bea
\frac{dn_{\bm k}}{d \tau}&=& i \omega_0 \epsilon 
\left( \tilde{m}_{\bm k} - \tilde{m}_{\bm k}^* \right)
+ g^2\frac{\sigma^2}{4\omega_0^2} \left(n_{\bm k}+\frac{1}{2} \right), 
\label{DRGneq}\\
\frac{d\tilde{m}_{\bm k}}{d \tau}&=& - i \omega_0 \epsilon 
\left( 2 \tilde{a}_{\bm k} \tilde{m}_{\bm k} + 2 n_{\bm k} + 1 \right) 
-g^2\frac{\sigma^2}{8\omega_0^2} \tilde{m}_{\bm k}.
\label{DRGmeq}
\eea

In the absence of the noise term these equations are easily solved and
the solution for the particle number with the initial conditions,
$n_{\bm k}(0)=m_{\bm k}(0)=0$, is found to be
\be
n_{\bm k}(t)= \frac{1}{1-\tilde{a}_{\bm k}^2} \sinh^2 (\gamma_{\bm k} t),
\ee
where the Floquet exponent is given by
\be
\gamma_{\bm k} = \epsilon \omega_0 \sqrt{1-\tilde{a}_{\bm k}^2}.
\ee
This coincides exactly with the well-known results for the parametric
resonance in the literature. 
Resonance occurs for the particle modes with $|\tilde{a}_{\bm k}| < 1$,
otherwise the particle number shows oscillation.

However it is explicitly seen that the noise term alters the nature of the
solutions of the equations (\ref{DRGneq}) and (\ref{DRGmeq}). 
We may evaluate the Floquet exponent by assuming the asymptotic form 
as $n_{\bm k}(t) \simeq A \exp (2 \gamma_{\bm k} t)$ and so on.
Then the exponent is given by the largest eigenvalue of the matrix
\be
\left(
\begin{array}{ccc}
2\Delta & 0 &  -2 \omega_0 \epsilon \\
0 & -\Delta  & 2\omega_0 \epsilon \tilde{a}_{\bm k} \\
-2 \omega_0 \epsilon & -2  \omega_0 \epsilon \tilde{a}_{\bm k}
& -\Delta
\end{array}
\right),
\label{the matrix}
\ee 
where $\Delta = g^2 \sigma^2/8\omega_0^2$.
For a small enough $\Delta$ the exponent may be evaluated
as 
\be
\gamma_{\bm k} = \epsilon \omega_0 \sqrt{1-\tilde{a}_{\bm k}^2}+
\frac{\sigma^2(1+2\tilde{a}_{\bm k}^2)}{32\omega_0^2(1-\tilde{a}_{\bm
k}^2)}\,
g^2+O(\Delta^2).
\label{Floquet exponent}
\ee
Thus it is found that the exponent is always enhanced: the second term
is positive for $|\tilde{a}_{\bm k}| < 1$.  Moreover even in the stable
region of the Mathieu equation, $|\tilde{a}_{\bm k}| > 1$, the Floquet
exponent has real positive part and the resonance occurs.  Thus we find
that the stable region disappear at least near the first resonance
peak. We will show the results by numerical analysis of
Eqs.~(\ref{DRGneq}) and (\ref{DRGmeq}) in section~5.

A comment is in order.  The Floquet exponent for $|\tilde{a}_{\bm k}| >
1$ is not the one given in Eq. (\ref{Floquet exponent}).  It is easy to
confirm that the largest eigenvalue of the traceless matrix in Eq.~(\ref{the
matrix}) is positive for $|\tilde{a}_{\bm k}| > 1$.

\section{Exact evolution equations}

The analysis by the DRG is based on the
perturbative expansion, although the infinite series of the leading
secular terms are summed up.  Therefore the results are not always
reliable for the broad resonance cases, $\epsilon \geq O(1)$, or for the
cases with large noise effects.  Moreover, the DRG is not suitable for the purpose of studying the particle
production for the wide range of the momentum modes simultaneously.  As
seen explicitly in the next section, the above DRG equation gives a good
result for the momentum only around the first resonance peak.

It was found in Ref.~\cite{HMNT} that the evolution
of the particle numbers can be evaluated exactly by applying a variation
of exact renormalization group equation in the absence of the
self-interactions.  The exact renormalization group equation is also
consistent with time evolution by the equation of motion. In this
section we apply this method to the present system with the noise effect
and we find exact evolution equations.

Before going into this discussion, let us remind the basic equations
in generic non-equilibrium field theories~\cite{Rammer:2007zz}.
First we introduce the statistical function $F_{\bm k}$ and 
the spectral function $\rho_{\bm k}$ as\footnote{In the thermal equilibrium
these functions are related to each other through the
fluctuation-dissipation relation. However these functions are
independent in generic non-equilibrium situations. }
\be
G_{\bm k}(t,t')
= F_{\bm k}(t,t') - \frac{i}{2}\rho_{\bm k}(t,t')\mbox{sgn}(t-t').
\ee
Therefore 
\bea
F_{\bm k}(t,t') &=& \frac{1}{2}\left[ G^>(t,t')+G^<(t,t') \right] 
= \frac{1}{2}\lan \left\{ \chi_{\bm k} (t), \chi_{\bm k} (t') \right\} \ran, \\
\rho_{\bm k} &=& i \left[ G^>(t,t') - G^<(t,t') \right] 
= i \lan \left[ \chi_{\bm k} (t), \chi_{\bm k} (t') \right] \ran.
\eea
It is noted that the relations 
$\rho_{\bm k}(t,t')|_{t=t'}=0$ and  $\partial_t \rho_{\bm k}(t,t')|_{t=t'}=1$ hold.

The particle number $n_{\bm k}(t)$ may be evaluated from the statistical
function as
\bea
n_{\bm k}(t) + \frac{1}{2}
&=&
\left.
\frac{1}{2\omega}\left( \partial_t \partial_{t'} + \omega^2 \right)
\lan \chi_{\bm k} (t) \chi_{\bm k} (t') \ran  
\right|_{t' \rightarrow t} \nn \\
&=& 
\left.
\frac{1}{2\omega} \left( \partial_t \partial_{t'} + \omega^2 \right)
F_{\bm k}(t, t')   
\right|_{t' \rightarrow t}.
\eea
Therefore we consider to evaluate the statistical functions.  
For a free particle system with a general density matrix $\rho(t_0)$, 
the statistical function is explicitly given by 
\bea
F_{\bm k}(t,t')&=&
\frac{1}{2 \omega} \Bigl\{
2\cos[\omega(t-t')]\left(n_{\bm k}(t_0)+\frac{1}{2} \right) \nn \\
& & \hspace{24pt} 
+ e^{-i \omega(t+t')}\tilde{m}_{\bm k}(t_0)
+ e^{i \omega(t+t')}\tilde{m}_{\bm k}^*(t_0).
\Bigr\},
\eea
while the spectral functions is given by Eq.~(22).

Now we introduce a cutoff to the interaction time. Explicitly
we define the two point function cutoff at $\tau$,
$G_{\bm k}^>(t,t')_{\tau}$, by
\be
G_{\bm k}^>(t,t')_{\tau} =
\int D\chi^+ D\chi^- \chi^+_{\bm k}(t) \chi^-_{\bm k}(t')~
e^{i S_{\rm eff}[\chi^+, \chi^-]}\,,
\ee
where
\bea
S_{\rm eff}&=&
\int_0 dt \sum_{\bm k}
\Bigl\{ \frac{1}{2}|\dot{\chi}^+_{\bm k}(t)|^2-
\frac{1}{2}\omega^2|\chi^+_{\bm k}(t)|^2
\Bigr\} 
-\Bigl\{ \frac{1}{2}|\dot{\chi}^-_{\bm k}(t)|^2-
\frac{1}{2}\omega^2|\chi^-_{\bm k}(t)|^2
\Bigr\} \nn \\
& &
- \int_0^{\tau} dt
\Bigl\{
\frac{1}{2}\epsilon f(t) \sum_{\bm k}
\left( |\chi^+_{\bm k}(t)|^2 -|\chi^-_{\bm k}(t)|^2 \right) \nn \\
& & 
\hspace{36pt}- i g^2 \frac{\sigma^2}{8}  \sum_{\bm k} \sum_{{\bm k}'} 
\left( \chi^+_{\bm k}(t)^2 - \chi_{\bm k}^- (t)^2 \right)
\left( \chi^+_{{\bm k}'}(t)^2 - \chi_{{\bm k}'}^- (t)^2 \right)
\Bigr\}.
\eea
Note that $G_{\bm k}^>(t,t')_{\tau}$ is independent of the cutoff time $\tau$
when $\tau > t, t'$. 
Therefore it is enough to consider the cases of $0< \tau < t,t'$.

Then we may deduce the cutoff dependence of the two-point function  as
\bea
i \partial_{\tau} G_{\bm k}^>(t,t')_{\tau} 
&=&
\int D\chi^+ D\chi^- \chi^+_{\bm k}(t) \chi^-_{\bm k}(t')~e^{i S_{\rm eff}[\chi^+, \chi^-]}\nn \\
& &
\times \Bigl\{
\frac{1}{2}\epsilon f(\tau) \sum_{{\bm k}'} 
\left(|\chi_{\bm k}^+(\tau)|^2 - |\chi_{\bm k}^-(\tau)|^2 \right)\nn \\
& &
- i g^2 \frac{\sigma^2}{8}  \sum_{\bm k'} \sum_{\bm k''} 
\left( \chi^+_{\bm k}(\tau)^2 - \chi_{\bm k}^- (\tau)^2 \right)
\left( \chi^+_{{\bm k}'}(\tau)^2 - \chi_{{\bm k}'}^- (\tau)^2 \right)
\Bigr\} \nn \\
&=&
\frac{1}{2}\epsilon f(\tau) \sum_{{\bm k}'} 
\Big\{ 
\lan \chi_{\bm k}(t) \chi_{\bm k}(t') \chi_{\bm k'}^2(\tau)\ran
- \lan  \chi_{\bm k'}^2(\tau) \chi_{\bm k}(t) \chi_{\bm k}(t') \ran
\Bigr\} \nn \\
& &
-ig^2\frac{\sigma^2}{8} 
 \sum_{\bm k'} \sum_{\bm k''}\Big\{ 
\lan \chi_{\bm k}(t) \chi_{\bm k}(t') \chi_{\bm k'}^2(\tau) \chi_{\bm k''}^2(\tau) \ran
+ \lan \chi_{\bm k'}^2(\tau) \chi_{\bm k''}^2(\tau) \chi_{\bm k}(t) \chi_{\bm k}(t') \ran 
\nn \\
& &
\hspace{48pt}
- \lan \chi_{\bm k'}^2(\tau) \chi_{\bm k}(t) \chi_{\bm k}(t')  \chi_{\bm k''}^2(\tau) \ran
- \lan  \chi_{\bm k''}^2(\tau) \chi_{\bm k}(t) \chi_{\bm k}(t') \chi_{\bm k'}^2(\tau) \ran 
\Big\}.
\eea
Here we note that these correlation functions can be rewritten in terms
of the commutators of the scalar fields, which is given by 
\bea
 [\chi_{\bm k}(t),  \chi_{\bm k'}(\tau)]
&=& \left( D^>_{\bm k}(t-\tau) - D^>_{\bm k}(\tau-t) \right) \delta_{{\bm k}{\bm k'}}\nn \\
&=& -i  \rho_{\bm k}(t-\tau)\delta_{{\bm k}{\bm k'}}.
\eea
Eventually the cutoff dependence of the two-point function is found to be
\bea
i \partial_{\tau} G_{\bm k}^>(t,t')_{\tau}
&=&
-i \epsilon f(\tau)\Big\{
G_{\bm k}^>(t,\tau)_{\tau} \rho_{\bm k}(t'-\tau)
+ G_{\bm k}^>(\tau,t')_{\tau}  \rho_{\bm k}(t-\tau)
\Bigr\} \nn \\
& &
+ i g^2 \sigma^2  G_{\bm k}^>(\tau,\tau)_{\tau}
 \rho_{\bm k}(t-\tau) \rho_{\bm k}(t'-\tau).
\eea
It should be noted that this evolution equation is exact.

If we decompose the cutoff two-point function as
\be
G_{\bm k}(t,t')_{\tau}=
F_{\bm k}(t,t')_{\tau}- \frac{i}{2}\rho_{\bm k}(t,t')_{\tau},
\ee
then it is immediately found that  the evolution equations for these
functions are given by
\bea
\partial_{\tau} F_{\bm k}(t,t')_{\tau}
&=&
-\frac{\epsilon f(\tau)}{\omega}
\left[
F_{\bm k}(t,\tau)_{\tau} \sin[\omega(t'-\tau)]+
F_{\bm k}(\tau,t')_{\tau} \sin[\omega(t-\tau)]
\right] \nn\\
& &
+ g^2\frac{\sigma^2}{\omega^2} F_{\bm k}(\tau,\tau)_{\tau}
\sin[\omega(t-\tau)]\sin[\omega(t'-\tau)], \label{ERGforF} \\
\partial_{\tau} \rho_{\bm k}(t,t')_{\tau}
&=&
-\frac{\epsilon f(\tau)}{\omega}
\left[
\rho_{\bm k}(t,\tau)_{\tau} \sin[\omega(t'-\tau)]+
\rho_{\bm k}(\tau,t')_{\tau} \sin[\omega(t-\tau)]
\right]. \label{ERGforrho}
\eea
We note 
\be
\rho_{\bm k}(t,0)_{\tau=0} = \frac{1}{\omega}\sin[\omega t],
\ee
since the interactions are totally switched off for $\tau=0$.
Therefore, we find from Eq.~(\ref{ERGforrho}) 
\be
\partial_{\tau} \rho_{\bm k}(t,t')_{\tau=0} = 0.
\ee
This means that the spectral function is always given by the free form
given by Eq.~(\ref{freerho}).\footnote{If thermalization occurs, then the
spectral function should evolve and approach to the form satisfying the
fluctuation-dissipation relation. Therefore this system does not show
thermalization. It seems to be caused by lack of self-interaction among
different momentum modes.}

Evolution of the particle number is derived from the Eq.~(\ref{ERGforF}). 
Since the system is free for $t, t' >\tau$, 
$ F_{\bm k}(t,t')_{\tau}$ is represented in terms of 
$n_{\bm k}(\tau)$ and $m_{\bm k}(\tau)$.
Therefore
\be
\partial_{\tau} F_{\bm k}(t,t')_{\tau}
= \frac{1}{2\omega}
\left\{
2 \cos[\omega (t-t')] \dot{n}_{\bm k}(\tau)
+ e^{-i\omega (t+t')} \dot{\tilde{m}}_{\bm k}(\tau)
+ e^{i\omega (t+t')} \dot{\tilde{m}}^*_{\bm k}(\tau)
\right\}.
\ee
Then, by noting that $\tilde{m}_{\bm
k}=e^{2i\omega\tau}m_{\bm k}$, the 
evolution equations for the particle number and the anomalous particle 
number may be obtained from Eq.~(\ref{ERGforF}) exactly as\footnote{As is discussed in Ref.~\cite{HMNT}, these equations may be
regarded as the exact renormalization equations.}

\bea
\frac{d n_{\bm k}}{d \tau} 
&=& \frac{i}{2\omega}\epsilon f(\tau)
\left( m_{\bm k} - m_{\bm k}^* \right) 
+ g^2 \frac{\sigma^2}{4 \omega^2}(2 n_{\bm k} + 1 + m_{\bm k} +  m_{\bm
k}^* ),
\label{eq:exact eq1}\\
\frac{d m_{\bm k}}{d \tau} 
&=& -2i \omega m_{\bm k} 
-\frac{i}{2\omega}\epsilon f(\tau)
\left( 2 n_{\bm k}+ 2m_{\bm k}+1 \right)
-g^2 \frac{\sigma^2}{4 \omega^2}(2 n_{\bm k} + 1 + m_{\bm k} +  m_{\bm k}^* ).
\label{eq:exact eq2}
\eea

We may solve these equations by perturbative expansion. Then we can reproduce
the DRG equations, after extracting the secular terms.

\section{Numerical analyses}

Here we study numerically how the noise affects the particle number spectra
by using 
the exact evolution equations in (\ref{eq:exact eq1}) and
(\ref{eq:exact eq2}) as well as the DRG equations in (\ref{DRGneq}) and
(\ref{DRGmeq}).

For a weak external force or for a small $\epsilon$, narrow resonance
peaks appear in the momentum spectra of the particle number, while, for
a strong external force, broad resonance peaks show up.  The situations
are called the narrow resonance and the broad resonance, respectively.
In the following subsections, we separately consider these two
situations.  Numerical calculations are performed by choosing the
parameters as $\omega_0=1$, $m_\chi=0$ and $\sigma=1$.

\subsection{Noise effect to narrow resonance case}

\begin{figure}[htbp]
\begin{center}
\includegraphics[width=80mm]{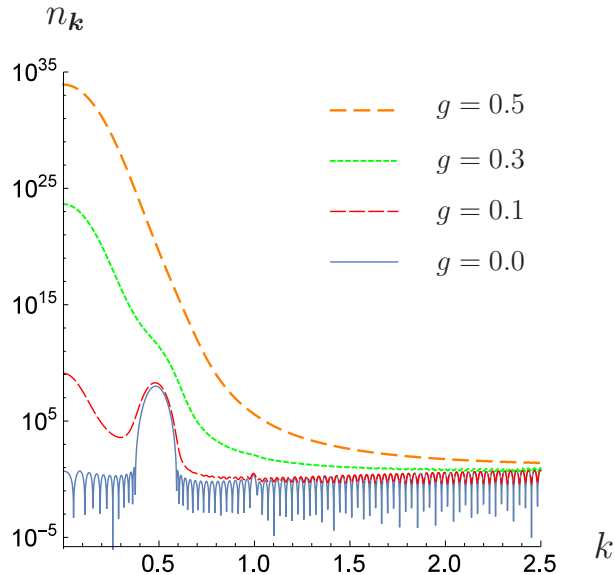}
\end{center}
 \caption{
The particle numbers $n_{\bm k}$ at $t=100$ obtained with the exact evolution equations with $\epsilon=0.1$.
The four lines are respectively for $g=0.0, 0.1, 0.3, 0.5$ from the bottom to the top. 
}
\label{fig:spectraERGn1}
\end{figure}

\begin{figure}[htbp]
\begin{center}
\includegraphics[width=80mm]{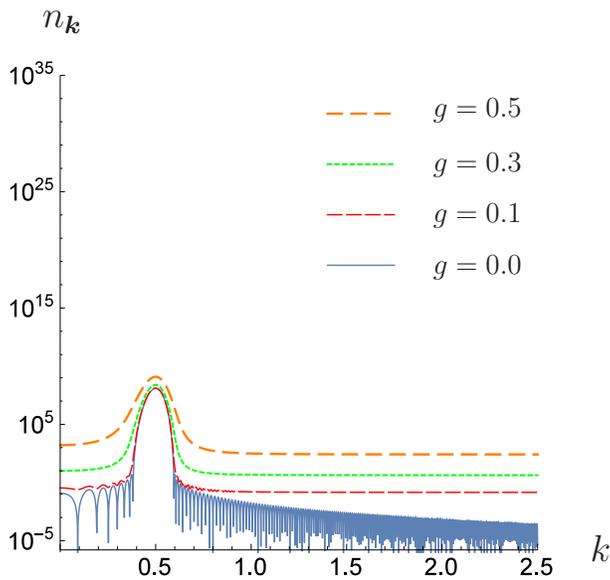}
\end{center}
 \caption{The particle numbers $n_{\bm k}$ at $t=100$ obtained by the DRG equations with $\epsilon=0.1$.
The four lines are respectively for $g=0.0, 0.1, 0.3, 0.5$ from the bottom to the top. 
}
\label{fig:spectraDRGn1}
\end{figure}

First, we consider a narrow resonance case with $\epsilon=0.1$.
Fig.~\ref{fig:spectraERGn1} and Fig.~\ref{fig:spectraDRGn1} show the
particle number spectra obtained with the exact evolution equations and
the DRG equations, respectively.  The particle numbers $n_{\bm k}(t)$ are plotted as functions of $k=|{\bm k}|$. 

The four lines in Fig.~\ref{fig:spectraERGn1} show the particle number
$n_{\bm k}(t)$ at $t=100$ under the noises with $g=0, 0.1, 0.3, 0.5$ from the
bottom to the top.  Clearly, the particle numbers are strongly enhanced
owing to the noise effect: the enhancement becomes more evident as the
noise increases; for a large enough noise, even the peaks of resonances
are not observed.

The line for $g=0$, i.e., without a noise, shows the resonant peak
around $k=1/2$.  The next peak is expected to appear around $k=1$ and we
observe a minor structure on the curve.  We find two oscillating regions
from $k=0$ to $k \sim 0.4$ and $k \sim 0.6$ to $k \sim 1.0$, where
$n_{\bm k}(t=100)$ oscillates as a function of $k$.  In the presence of
a noise, the particle number increases drastically and the oscillating
regions diminish.  This result agrees with the theorem given in
Ref.~\cite{Zanchin:1997gf}, which implies that the generalized Floquet
exponent in the presence of a noise is strictly positive and the
particle production rate is enhanced irrespectively of the particle mode
$k$.  A brief description of the theorem is given in the Appendix A.

Fig.~\ref{fig:spectraDRGn1} shows the particle numbers evaluated by the
DRG equations with the same parameter $\epsilon=0.1$.  The particle
number increases in the presence of a noise and this happens even for
oscillating regions.  However, the DRG equation fails to capture the
drastic enhancement as observed in Fig.~\ref{fig:spectraERGn1}.

In the DRG equations, we used a perturbative expansion at the center of
the momentum mode $k=0.5$.  Therefore, in the absence of a noise, we
find the same first peak both in Fig.~\ref{fig:spectraERGn1} and
Fig.~\ref{fig:spectraDRGn1} and, around this peak or $k \sim 1/2$, the
DRG equations reproduced $n_{\bm k}$ similar to the exact evolution equations.
The second resonant peak in Fig.~\ref{fig:spectraERGn1} is absent in
Fig.~\ref{fig:spectraDRGn1}.  This is due to the perturbative expansion
mentioned above: in order to have the second peak, we need to make an
expansion around $k=1$.  From a comparison of two figures around $k=1/2$, we also
notice that the particle numbers are underestimated in the DRG equations
in the presence of noises.

Next, we consider the time evolution of particle numbers
for particular momentum modes both in the resonance band and the
non-resonance band. 

\begin{figure}[htbp]
\begin{center}
\includegraphics[width=80mm]{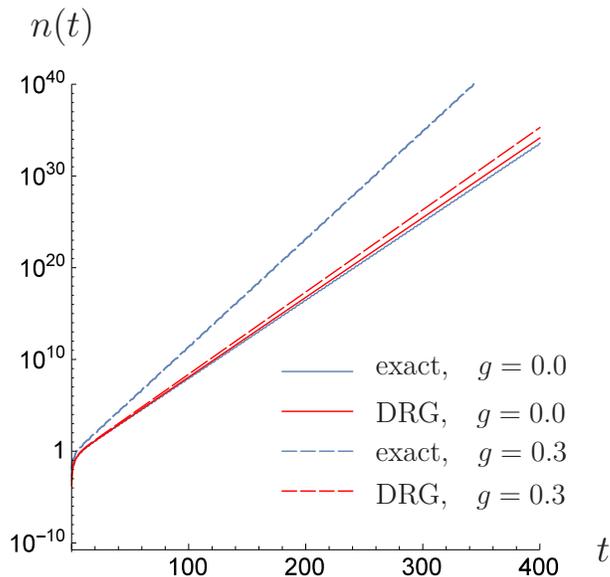}
\end{center}
\caption{(a) $n_{\bm k}(t)$ with $k=0.5$ in the narrow resonance with $\epsilon=0.1$.  The results with the exact evolution equations (the blue lines) and the DRG equations (the red lines) for
 $g=0.0, ~0.3$. }
\label{fig:exact evolution equation&DRG1}
\end{figure}

Fig.~\ref{fig:exact evolution equation&DRG1} shows $n_{\bm k}(t)$ for the mode
with $k=0.5$ in the resonance band.  The results from the exact
evolutions and the DRG equations are plotted with the blue and red
lines, respectively, for $g=0,~0.3$, with and without a noise.  In the
absence of a noise, both equations produce almost the same result.  However, for
both set of equations, a noise gives different results for the particle
production.  The enhancement is underestimated in the DRG equations.
\begin{figure}[htbp]
\begin{center}
\includegraphics[width=80mm]{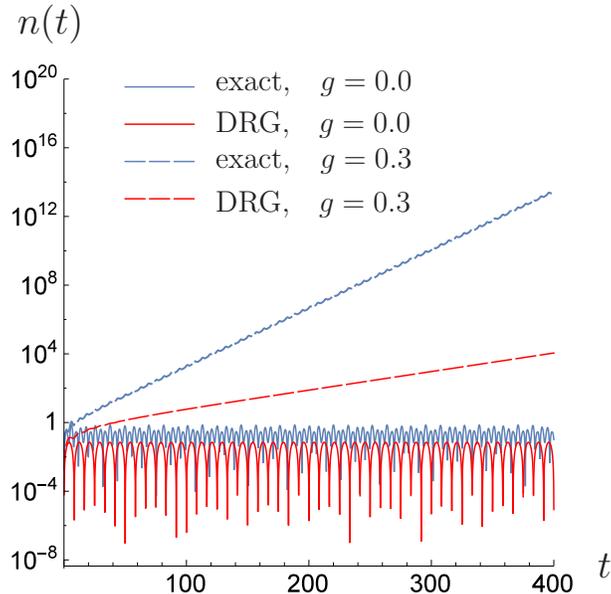}
\end{center}
\caption{(b) $n_{\bm k}(t)$ with $k=0.8$ in the narrow resonance with
 $\epsilon=0.1$.  The results with the exact evolution equations (the
 blue lines) and the DRG equations (the red lines) for 
 $g=0, ~0.3$.  The results of two methods are indistinguishable for
 $g=0$~.} 
\label{fig:exact evolution equation&DRG2}
\end{figure}

The results for the mode in the oscillating region with $k=0.8$ are
shown in Fig.~\ref{fig:exact evolution equation&DRG2}.  We observe the
expected behaviors: $n_{\bm k}(t)$ oscillates in the absence of noise
($g=0$), while it resonates and increases in the presence of noise
($g=0.3$).  For this mode, the DRG equations overestimate the particle
production.

\subsection{Noise effect to broad resonance case}

Taking the parameter $\epsilon=0.8$, we numerically studied the broad
resonance case: the same set of simulations are performed to produce
figures corresponding to Fig.~1 to Fig.~4, by using both the exact
evolution equations and the DRG equations.  The latter equations
however, are derived with a perturbative expansion and for a large
$\epsilon$ their results are not expected to be so reliable.  

The generic features of the noise effect observed for the narrow
resonance case is evident for the broad resonance case as well.  The
particle production rate rises 
irrespectively to the particle momentum $k$ and the effect wipes out the
resonance structure present without a noise.
One can observe this effect in Fig.~5 that plots the particle numbers
$n_{\bm k}(t)$ at $t=100$ for the values of $g=0, ~0.1, ~0.3, ~0.5$.  
\begin{figure}[htbp]
\begin{center}
\includegraphics[width=80mm]{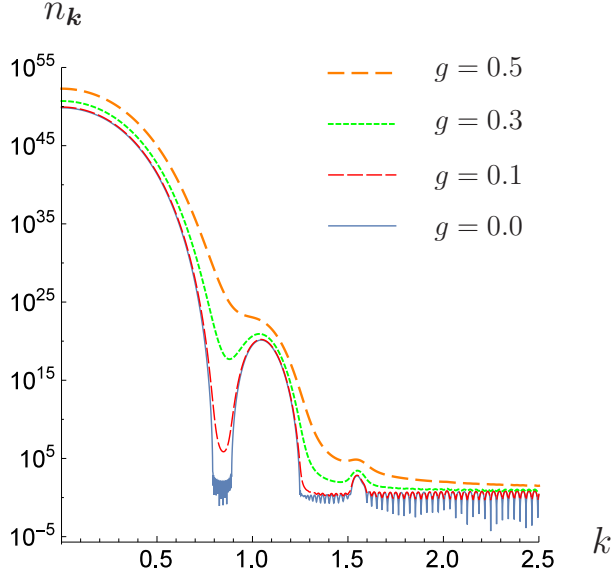}
\end{center}
 \caption{The particle numbers with $\epsilon=0.8$ corresponding to a
 broad resonance band:  $n_{\bm k}$ at $t=100$ obtained by the exact
 evolution equations for $g=0, ~0.1, ~0.3, ~0.5$. 
}
\label{fig:spectraERG1}
\end{figure}

\section{Summary and further questions}

In this article we discussed a scalar field theory with
the non-perturbative methods in order to consider the expected explosive
particle production at the end of the inflation epoch.  The scalar field
is coupled to an external force, the oscillating inflaton field, and
exposed to a Gaussian white noise.  

In Ref.~\cite{Zanchin:1997gf} it has been shown that the particle
production is enhanced by the noise effect.  However a concrete
method to evaluate the production rate has been missing.  In this paper,
we derived the evolution equations for the particle numbers in the
non-equilibrium quantum field theory and reported numerical results.

The correlation functions for non-equilibrium quantum fields  
are represented by path integrals of the Schwinger-Keldish formalism. 
After taking average over the Gaussian noise, the noise effect is 
incorporated as self-interaction terms with a pure imaginary
coupling constant. Therefore we discussed the particle numbers or the
two point functions with such interaction terms and examined the
parametric resonance phenomena.

Firstly, we obtained the DRG equations for the particle numbers based on
the non-equilibrium perturbative expansion at the leading order.  Then
it is explicitly shown that the production rate is enhanced in the
presence of the noise, as long as the expansion parameters are small
enough.

Secondly, we derived the evolution equations for the particle numbers
by introducing a cutoff to the interaction time and by evaluating the
response under a shift of the cutoff time.  The evolution
equations may be regarded as the non-equilibrium functional
renormalization group equations.
Surprisingly, the evolution equations for the particle numbers turned
out to be exact: owing to the special form of the noise interaction
terms, there was no need to use any approximation to write down the
evolutions equations.

In the last section we presented the various numerical results obtained
with the DRG and evolution equations for the particle numbers.  We examined
both the narrow resonance case and the broad resonance case. It is
clearly seen that the band structure in the momentum spectra of the
particle numbers disappears due to the noise effect.  The total
production rate of particles is also found to be enhanced.  The DRG
equations are expected to be reliable only when both the noise and the
external force are small.  Our numerical results are consistent with
this.

In Ref.~\cite{Zanchin:1998fj}, the authors extended their analysis to
the cases of spatially inhomogeneous noise and claimed that the particle
production rate is enhanced further compared to the homogeneous noise
case.  It would be interesting to extend our present analyses for the
case of inhomogeneous noise.  Furthermore, it is important to evaluate
the noise effect and the particle production rate in realistic inflation
scenarios.  We leave these questions for future studies.

\section*{Acknowledgements}

A.~M. is grateful to T.~Takahashi and K.~Itoh for their supports
in the final stage of writing up this paper.

\appendix

\section{Mathieu equation with noise and the Floquet exponent}

In this Appendix we consider a simplified differential equation given by
\be
\ddot{\chi} + (\omega^2 + p(t) + q(t))\chi = 0,  
\label{Mathieweq}
\ee
where $p(t)$ is a periodic function with a period $T$ and $q(t)$ denotes 
a noise function. Explicitly, we use
$p(t) =  2 \epsilon \cos t$ and, therefore $T=2 \pi$.
We can choose the two independent solutions $\chi_1(t)$ and $\chi_2(t)$
so that $\chi_1(0)=1, \dot{\chi}_1(0)=0$, $\chi_2(0)=0, \dot{\chi}_2(0)=1$.
Then the matrix
\be
\Phi_q(t,0)
=\left(
\begin{array}{cc}
\chi_1(t) & \chi_2(t) \\
\dot{\chi}_1(t) & \dot{\chi}_2(t)
\end{array}
\right)
\ee
gives the fundamental solution (or transfer) matrix of this equation.    
 
The fundamental solution satisfies the equation
\be
\dot{\Phi}_q = M(q(t),t) \Phi_q,
\ee
with the initial conditions $\Phi_q(0,0)=I$. Here $M(q(t), t)$ is the
matrix
\be
M(q(t),t)
=
\left(
\begin{array}{cc}
0 & 1 \\
-[\omega^2 + p(t) +q(t)] & 0
\end{array}
\right).
\ee
It is noted that the determinant of $\Phi_q(t)$ is the Wronskian of 
the differential equation (\ref{Mathieweq}),
which is a constant. Therefore $\mbox{det} \Phi_q(t) = 1$, {\it i.e.} 
$\Phi_q(t)$ is a matrix of SL$(2,{\bm R})$.

In the absence of noise the fundamental solution  is given in the form
\be
\Phi_0(t,0) = P_0(t)e^{Ct},
\ee 
where $P_0(t)$ is a periodic matrix function with period $T$ and $C$ is a constant matrix.
This is well-known as the Floquet theorem or the Bloch theorem.
The largest real part of the eigenvalues of $C$, $\mu(0)$, is called the Floquet exponent.
The matrix $\exp CT$ belongs to SL$(2,{\bm R})$, since $\Phi_0(T,0)=\exp CT$.  
Therefore we find  $\mu(0) \geq 0$. Resonance occurs when $\mu(0) >0$, while $\mu(0)=0$ 
in the stable region.

It has been shown in Ref.~\cite{Zanchin:1997gf} that the Floquet
exponent is always enhanced by the noise $q$.  The authors defined the
generalized Floquet exponent $\mu(q)$ by \be \mu(q) = \lim_{N
\rightarrow \infty} \frac{1}{NT} \log \lVert \prod_{j=1}^{N} \Phi_q(jT,
(j-1)T) \rVert, \ee where $\lVert ~~ \rVert$ denotes the matrix norm.
In order to see the noise effect to the Floquet exponent they introduced
the reduced transfer matrix $\Psi_q(t,0)$ by \be
\Phi_q(t,0)=\Phi_0(t,0)\Psi_q(t,0)=P_0(t)e^{Ct}\Psi_q(t,0).  \ee It is
noted that $\Psi_q(t,0)$ is also a SL$(2,{\bm R})$ matrix.

On the other hand, the Furstenberg's theorem tells us that independent
identically distributed random matrices $\Psi_j$ of SL$(2,{\bm R})$
enjoy \be \lim_{N \rightarrow \infty} \frac{1}{NT} \log \lVert
\prod_{j=1}^{N} \Psi_j \rVert \equiv \lambda > 0.  \ee Then we may apply
the Furstenberg's theorem to the reduced transfer matrix decomposed as
\be \Psi_q(NT,0)= \prod_{j=1}^N \Psi_q(jT,(j-1)T), \ee as long as the
transfer matrices $\Psi_q(jT,(j-1)T)$ are given by independent and
identically distributed random matrices.  Thus they showed
\be
\mu(q) = \mu(0) + \lambda > \mu(0).
\label{the theorem}
\ee
This result implies that the particle production rate is enhanced in the
presence of noise irrespectively of the particle mode.

\section{Numerical analysis of the noise effect in the Mathiew equation}

Here we would like to confirm the implication of
Eq. (\ref{the theorem}) by numerically calculating the particle number
 \be
n(t)= \frac{1}{2 \omega}(\dot{\chi}^2 + \omega^2 \chi^2)
\label{particle number}
\ee
from the classical equation (\ref{Mathieweq}).  $n(t)$ in Eq.~(\ref{particle
number}) may be regarded as the classical energy.

The technically important point turns out to be how we introduce a noise function.
Naively we may divide the time region into intervals 
and approximate the noise function as a sequence of independent constants assigned to
intervals~\cite{Zanchin:1997gf}. However it is found that the Floquet
exponent is not always enhanced in the resonant band. Rather the
exponent is reduced at the resonant peak in the presence of
noise.
%
This seems to be contradicting with
the above proof of the universal enhancement of the Floquet exponent.

Here we propose a different realization of the noise function.  
We represent the noise function as a superposition given by
\be
\nu(t) = \frac{a_0^{(j)}}{2} + \sum_{n=1}^N (a_n^{(j)} \cos n t + b_n^{(j)} \sin n t).
\label{noiseFourier}
\ee
for each interval of $(j-1)T < t < jT$. The coefficients are given by
random variables and are independent for different intervals;
\be
\langle a_n^{(j)} a_m^{(j')} \rangle = \delta_{n,m} \delta_{j, j'}.  
\ee
Then the white noise property, $\langle \nu(t) \nu(t') \rangle =
\delta(t-t')$ is maintained.

\begin{figure}[htbp]
\begin{center}
\includegraphics[width=80mm]{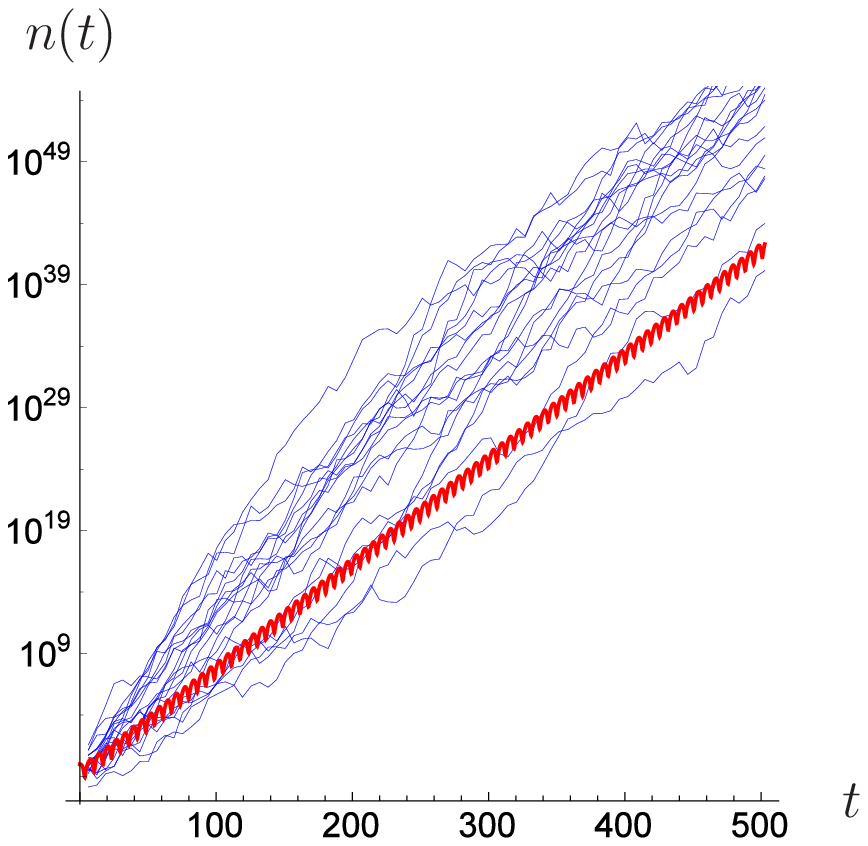}
\end{center}
\caption{$n(t)$ obtained by solving Eq.~(\ref{Mathieweq}) with
$\epsilon=0.1$ and $\omega^2=1/4$ (the first resonance peak). The noise function
$q(t) = 3 \nu(t)$, where $\nu(t)$ is given by Eq.~(\ref{noiseFourier})
with $N=5$.  The red line shows $n(t)$ in the absence of a noise, while
the blue lines are $n(t)$ obtained with various noise functions.  }
\label{fig:Mathieuwithnoise1}
\end{figure}

In Fig.~\ref{fig:Mathieuwithnoise1}, evolutions of the particle number
$n(t)$ are shown in a logarithmic plot in the narrow resonance
case. Explicitly the parameters of Eq.~(\ref{Mathieweq}) are set as
$\omega^2=1/4$ and $\epsilon=0.1$, which correspond to the first resonance
peak. The noise function is set to be $q(t) = 3 \nu(t)$, where $\nu(t)$ 
is given by Eq.~(\ref{noiseFourier}) with $N=5$. In
Fig.~\ref{fig:Mathieuwithnoise1} the red line shows the particle number
in the absence of the noise, and other blue lines show those obtained
with various noise functions.

The particle numbers are found to grow exponentially. The Floquet
exponents can be read off from the  gradients of these lines.
The generalized Floquet exponent $\mu(q)$ is given by the average of the
gradients of blue lines. It is clearly seen that the Floquet exponents
are enhanced in the presence of noise. It is noted that only the lowest
frequency modes of the noise function are effective, though we performed
these calculations for $N=5$. So it is thought that enhancement occurs
because such a mode acts as a fluctuation to the amplitude of external
force or $\epsilon$.   
\begin{figure}[htbp]
\begin{center}
\includegraphics[width=80mm]{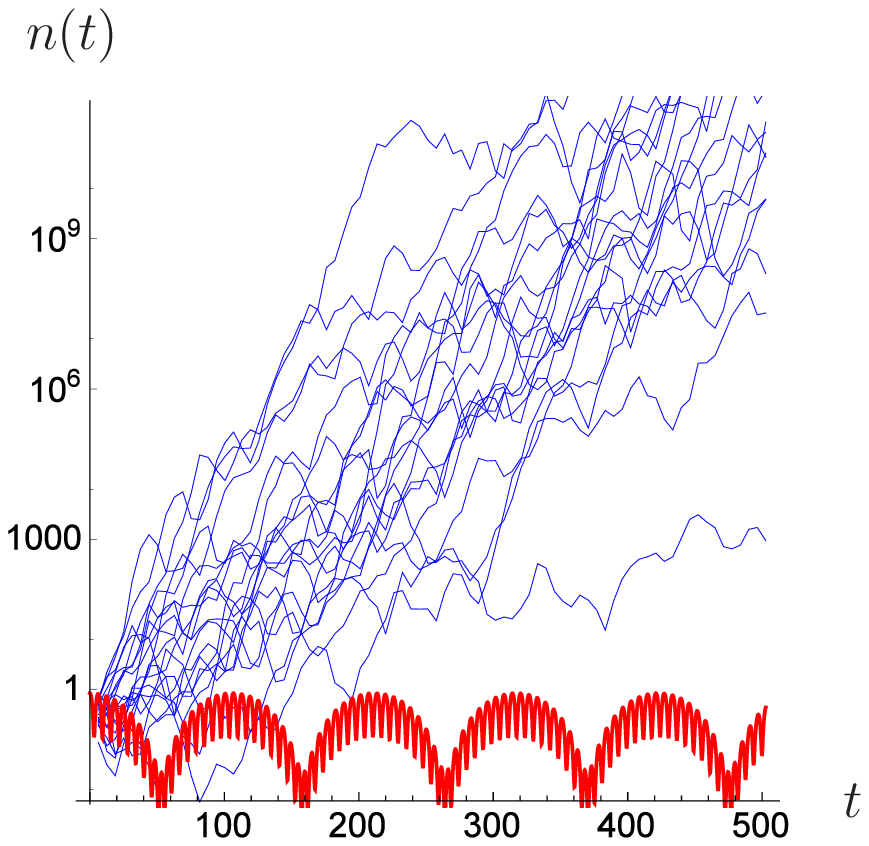}
\end{center}
 \caption{$n(t)$ obtained by solving Eq.~(\ref{Mathieweq}) with
$\epsilon=0.1$ and $\omega^2=1/4+1/10$ (in a slightly stable region off
the first resonance peak). The noise function $q(t) = \nu(t)$, where
$\nu(t)$ is given by Eq.~(\ref{noiseFourier}) with $N=5$.  The red line
shows $n(t)$ in the absence of a noise, while the blue lines are $n(t)$
obtained with various noise functions.  } \label{fig:Mathieuwithnoise2}
\end{figure}

We also examined the case in a stable region of the Mathieu equation,
by choosing the parameters as $\epsilon=0.1$ and
$\omega^2=1/4+1/10$.  While the particle number obtained from a solution of
the Mathieu equation is shown by the red line in
Fig.~\ref{fig:Mathieuwithnoise2}, it oscillates  between $(0, 1)$. Then
the Floquet exponent is given by a pure imaginary number.

The blue lines in Fig.~\ref{fig:Mathieuwithnoise2} show the particle
numbers obtained in the presence of the noise function $q(t) = \nu(t)$,
where $\nu(t)$ is given by Eq.~(\ref{noiseFourier}) with $N=5$. It
is seen that the particle numbers grow exponentially.

Our numerical results show that the particle number defined in
Eq. ~(\ref{particle number}) increases exponentially in the presence of a
noise, irrespectively whether the parameters are stable or unstable
regions for the Mathieu equation.  Our result is consistent with the
analytical result Eq.~(\ref{the theorem}) explained in Appendix A.


\end{document}